\documentclass[%
 amsmath,amssymb,
 aps,
twocolumn, 
]{revtex4-2}

\usepackage{graphicx}
\usepackage{dcolumn}
\usepackage{bm}

\usepackage{siunitx}
\usepackage{cleveref}
\crefname{figure}{Fig.}{Figs.} 

\usepackage{braket}
\usepackage{xcolor}
\definecolor{purple}{rgb}{0.60, 0.0, 0.60}

\begin{document}

\preprint{APS/Phys. Rev. Appl.}

\title{Laser Frequency Stabilization Using Light Shift in Compact Atomic Clocks} 

\author{Claudio E. Calosso}
\author{Michele Gozzelino} \email{m.gozzelino@inrim.it}
\author{Filippo Levi}%
\author{Salvatore Micalizio}
\affiliation{%
 Istituto Nazionale di Ricerca Metrologica, INRIM, Quantum Metrology and Nanotechnologies Division, Strada delle Cacce 91,  Torino, 10135, Italy
}%

\date{\today}

\begin{abstract}
This paper describes the Light-Shift Laser-Lock (LSLL) technique, a novel method intended for compact atomic clocks that greatly simplifies the laser setup by stabilizing the pumping-laser frequency to the atoms involved in the clock, without the need of an external reference. 
By alternating two clock sequences with different light shifts, the method estimates and cancels out a controlled amount of induced light shift, acting on the laser frequency.
The LSLL technique is compatible with state-of-the-art 3-level clocks and was demonstrated with FPGA-based electronics on a pulsed-optically-pumped (POP) vapor-cell clock developed at INRIM. The results have shown that the LSLL technique operates robustly, having a capture range of gigahertz without significantly compromising clock stability. In our tests, the clock exhibited a white frequency noise of $\SI{3.2e-13} \tau^{-1/2}$ for averaging time up to \SI{4000}{s}, reaching a floor below \SI{1e-14} up to \SI{100000}{s}. 
These performance levels meet the requirements of future Global Navigation Satellite Systems (GNSS) on-board clocks, and offer the added benefits of a reduced clock footprint, as well as increased reliability and robustness.
\end{abstract}

\maketitle


\section{Introduction}
\label{sec:intro}
Laser stabilization is a key tool in atomic and molecular physics experiments, as well as in quantum-communication platforms \cite{Demtroder2002}. Often, the frequency reference is a high-finesse Fabry-Perot cavity or an interferometer \cite{Robinson2019, Milner2019, Kong2015}. When the quality factor of the frequency reference is high, the stabilization can also bring to line narrowing. However, being based on physical objects, these frequency references can experience drifts and non-stationary behaviors, which may not always be suitable for applications requiring long-term stability or an absolute frequency reference \cite{Milner2019}. In such cases, stabilizing the laser frequency to an atomic transition is often the most effective approach. \cite{ Preuschoff2018, Corwin1998, Wieman1976, Strangfeld2022}. 

In laser-pumped atomic clocks, laser light is utilized to prepare the atoms in the desired ground state and to read out the atomic population. 
Active stabilization of the laser frequency is essential in these cases as well, since frequency-dependent light-shift effects can introduce instability to the clock transition \cite{Vanier1989, Vanier2007}. 
The degree of instability induced by light-shift  depends on the physical principles of the clock, often requiring state-of-the-art laser stabilization \cite{Bandi2014,Yun2017,McGuyer2009}. This is also the case with coherent-population-trapping (CPT) clocks, where different techniques have been proposed to mitigate light-shift effects \cite{Hafiz2018, Hafiz2020, Yudin2018a, Shah2006}. Usually, the laser frequency is stabilized on a dedicated external spectroscopy setup. However, adding a dedicated branch for probing an external reference can increase the complexity, size, weight of the laser system, which is not desirable for real-world applications \cite{Micalizio2021}. 

In this paper, we turn the concept around and show that actually the light-shift can be a convenient tool to frequency-stabilize the laser leading to the development of a more compact and robust frequency standard. 
The idea of using the light-shift for stabilization was first proposed a while back by Arditi and Picqu\'e, who compared two clocks: one with enhanced light shift and another with regular light shift, serving as a reference \cite{Arditi1975}. 
Our work demonstrates that a single clock alternating between high- and low- light-shift modes can replace the need for an external atomic clock, introducing a new technique named ``light-shift laser lock" (LSLL).

Since the light-shift curve has a dispersive profile whose width is proportional to the line strength, the LSLL technique is particularly suitable to stabilize lasers resonant to strong allowed transitions, 
commonly used for optical pumping and read-out. This characteristics naturally provides a wide capture range, making this method appealing for automatic re-lock procedures and unmanned operation. 

The technique takes advantage of the same atomic sample used in the clock operation, offering a dual benefit. First, the atoms are usually in a very well controlled environment, especially in terms of electromagnetic fields and temperature, a desirable  condition for best frequency stability in the long term. Second, being the atomic signal already available, the technique can be implemented with minimal modifications to the clock hardware, mainly relying on firmware or software implementation. This peculiarity makes the method suitable for an easy upgrade in existing systems, provided that they are based on a digital architecture.

Undoubtedly, there are close analogies between the (symmetric) auto-balanced Ramsey method \cite{Sanner2018, Hafiz2018} and other advanced protocols that detect and correct for the light-shift effect \cite{Hafiz2020, Yudin2018a}.
In fact, in all these cases, the light-shift is estimated by alternating low and high light-shift clock sequences. However, while those techniques compensate for the light-shift by acting on the microwave synthesizer, leaving the atoms in the same perturbed conditions, this technique directly acts on the laser frequency to minimize its influence, thereby leaving the atomic sample in a less perturbed state.

We implement the technique in a Pulsed Optically Pumped (POP) clock, where we stabilized the pumping laser without the use of an external frequency reference, and demonstrate state-of-the-art clock stability. This is a major step in increasing the robustness of the POP technology, especially for space applications \cite{Arpesi2019}. We note also that the LSLL technique does not need frequency modulation, further simplifying the laser setup and preserving the frequency-noise characteristic of the free-running laser, avoiding possibly detrimental amplitude modulation and aliasing effects \cite{Calosso2020}.

In conclusion, we point out that all these benefits come at the cost of a more complex clock sequence, which can pose a challenge for analog implementations or laboratory instrumentation-based setups. However, as mentioned earlier, this technique lends itself readily to digital architectures, such as those described in \cite{MTT17}.

\section{Technique}
\label{sec:technique}
\begin{figure}[ht]
    \centering
    \includegraphics[width=\columnwidth]{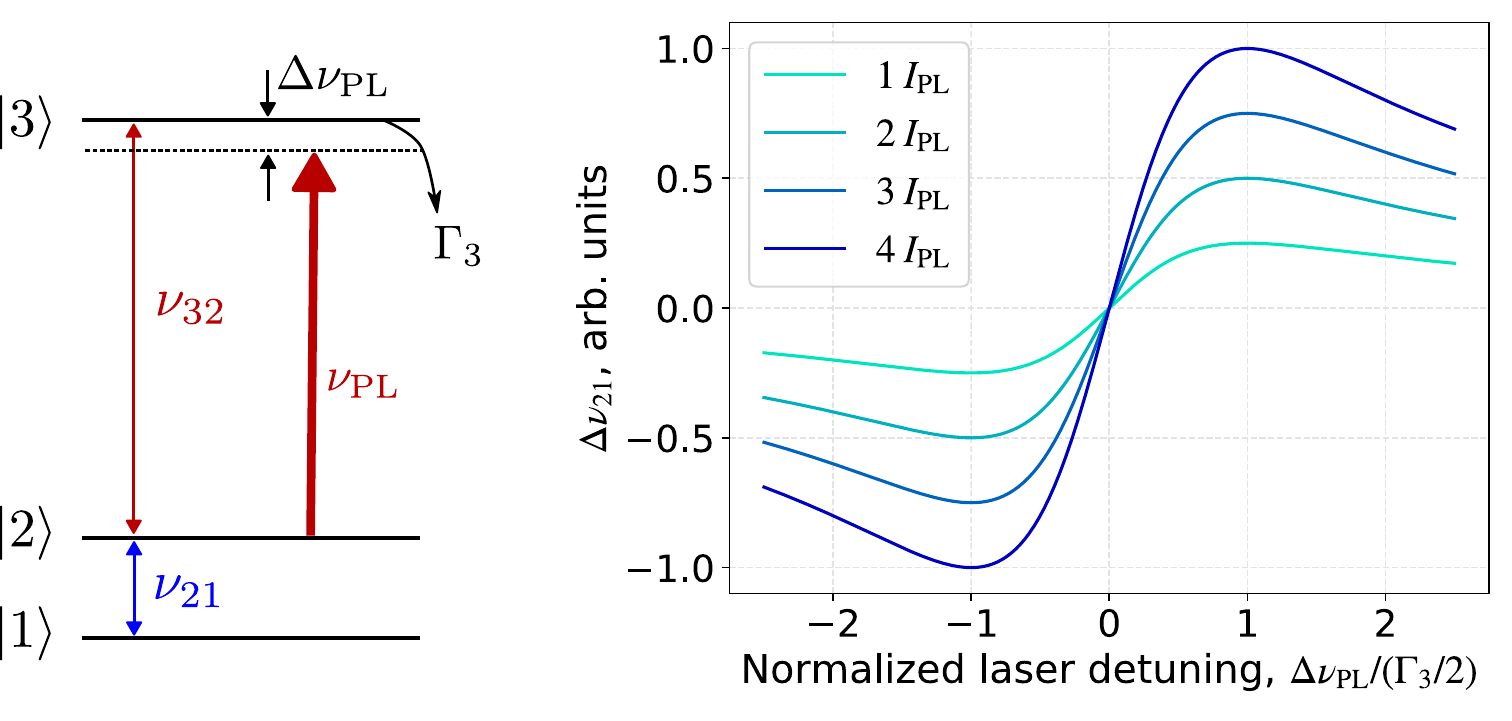}
    \caption{Left: simplified 3-level scheme common to 
     clocks based on a $\Lambda$-scheme. Same concepts applies to V-schemes. $\Delta\nu_{\text{PL}}$ is the laser frequency detuning from the transition linking one of the two ground states to the excited state. $\Gamma_3$ is the relaxation rate of the excited state. Right: light-shift affecting the clock transition ($\Delta\nu_{21}$) when the laser is nearly resonant with one of the two ground states as a function of the laser intensity during ($I_{\text{PL}}$, see text).
    }
    \label{fig:Fig1}
\end{figure}

Many compact clocks share a level structure sketched in \cref{fig:Fig1}a. The clock transition $\nu_{21}=\nu_2-\nu_1$ lies between two ground states $\ket{1}$ and $\ket{2}$ and a third state is used to perform optical pumping, thus creating a population imbalance \cite{Vanier1989}.  The presence of the optical field generates a frequency shift on the two clock states \cite{cohen-tannoudji2011}. In the weak-coupling regime ($\Omega<<\Gamma$), and with the laser resonant with one of the two ground states, we can approximate the light shift on the clock transition as: 

\begin{equation}
    \Delta\nu_{21} \simeq \frac{(-1)^i}{4} \frac{\Delta\nu_{\text{PL}}}{\Delta\nu_{\text{PL}}^2 + (\Gamma_3/2)^2} \Omega_{i3}^2
\end{equation}
where $\Delta\nu_{\text{PL}} = \nu_{\text{PL}} - \nu_{i3}$ is the laser detuning from the atomic transition linking the ground state (either $i=1$ or $i=2$) to the excited state $\ket{3}$ and ${\Omega_{i3}=\bra{i}e\mathbf{r}\cdot \mathbf{E}\ket{3} /h}$ is the associated Rabi frequency for a laser field vector $\mathbf{E}$. $\Gamma_3$ is the relaxation rate of the level $\ket{3}$. What is relevant for the present discussion are the following features: close to resonance, the light shift is linear with respect to the laser detuning. The magnitude of the shift is proportional to the square of the Rabi frequency and thus to the intensity of the perturbing laser ($I_{\text{PL}}$). Finally, the width of the dispersive curve is proportional to the relaxation rate $\Gamma_3$.

Generally, at resonance ($\Delta\nu_{\text{PL}} \ll \Gamma_3)$, the clock frequency is characterized by a sensitivity to the pumping laser frequency expressed as ${ \beta=\Delta \mathsf{y}/\Delta\nu_{\text{PL}} }$, where $\mathsf{y}$ is the clock fractional frequency fluctuation. The contribution to the clock instability will therefore be $ {\sigma_\mathsf{y}(\tau) = \beta \, \sigma_{\nu_{\text{PL}}}(\tau) }$, where $\tau$ is the averaging time. 

\begin{figure*}[!ht]
    \centering
    \includegraphics[width=0.8\textwidth]{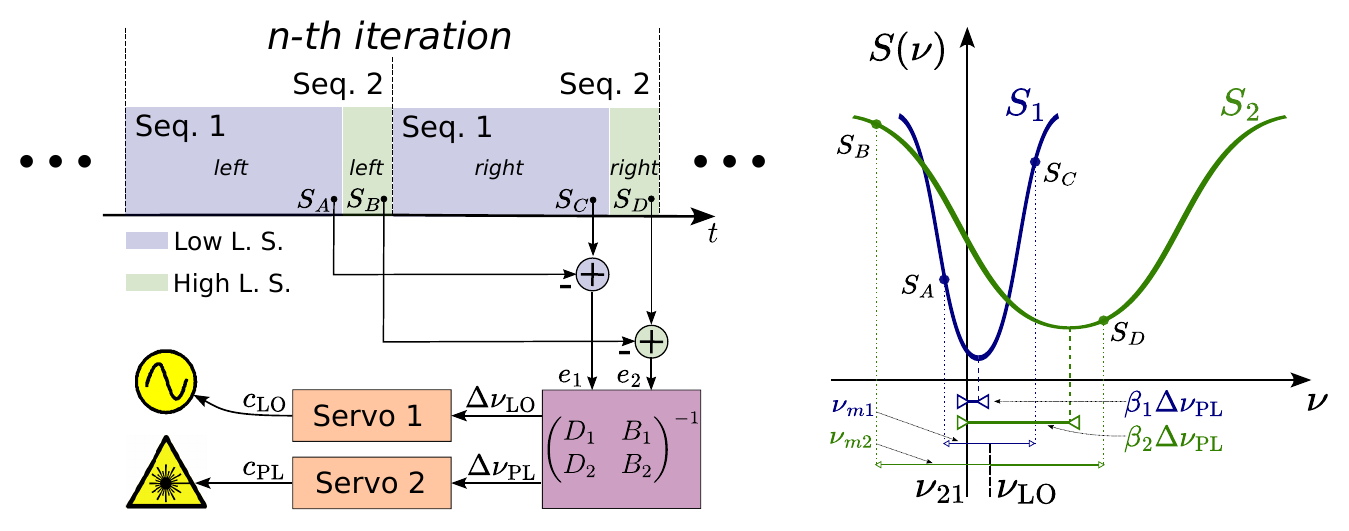}
    \caption{On the left, the operation principle of LSLL technique: the four atomic signals are employed to compute the error signals of the two sequences, which depend on both LO and PL detunings. The latters can be retrieved thanks the inverse of the sensitivity matrix and utilized by dedicated servos to correct both  the LO and PL frequencies.
    On the right, it can be observed how the spectroscopies, which are related to the two sequences, are differently affected by light shift, which is in turn proportional to the laser detuning. Additionally, the LO interrogation frequency is shown with their respective modulations that result in the four atomic signals used by the technique. By enabling the servos, the interrogating LO and the two curves align themselves with the unperturbed frequency, as aligning the PL cancels both light shifts. }
   
    \label{fig:scheme}
\end{figure*}

A common method used to maintain the laser in resonance and reduce the conversion of laser frequency to clock frequency is to stabilize the frequency of the pumping laser to an external reference. Alternatively, we can consider using the dispersive curve of the light shift as a frequency discriminator for this stabilization, as long as we have a suitable method for directly measuring the light shift. Experimentally, this is done by alternating two clock sequences: one for which the light-shift is minimal, and another in which the effect is enhanced by inducing a controlled light-shift, for example by sending a known amount of leaking optical radiation during the probing of the clock transition. The two clock sequences are labelled ``Seq. 1'' and ``Seq. 2'' in \cref{fig:scheme}. 

The error signals of the two clock sequences depend, although to different extents, on both the local oscillator (LO) frequency and the pumping laser (PL) frequency, as reported in the following equation:
\begin{align}
    \label{eq:system}
    \begin{pmatrix}
        e_1 \\ e_2
    \end{pmatrix} &= \begin{pmatrix}
        D_1 & B_1 \\
        D_2 & B_2 
    \end{pmatrix} \begin{pmatrix}
        \Delta\nu_{\text{LO}}\\
        \Delta\nu_{\text{PL}}
    \end{pmatrix}
\end{align}

where $\Delta\nu_{\text{LO}}= \nu_{\text{LO}}- \nu_{21}$ is the detuning of the LO frequency $\nu_{\text{LO}}$ from the atomic transition frequency estimated with the first clock loop ($\nu_{21}$) and $\Delta\nu_{\text{PL}}$ the detuning of the laser frequency $\nu_{\text{PL}}$ with respect to the frequency $\nu_{32}$. The coefficients $B_i$ are of course proportional to $\beta$,  whereas $D_i$ are inversely proportional to the atomic quality factor of the clock transition in the two clock sequences. $B_2$ can be purposely magnified by setting a controlled amount of laser intensity $I_{\text{PL}}$ during the clock interrogation of sequence 2.  

By inverting the $2	\times 2$ matrix presented in \cref{eq:system}, both $\Delta\nu_{\text{LO}}$ and $\Delta\nu_{\text{PL}}$ can be directly obtained and fed to two dedicated controllers that calculate the corrections $c_{\text{LO}}$ and $c_{\text{PL}}$ for the LO and PL frequencies respectively. $\Delta\nu_{\text{LO}}$ is used, as usual, to stabilize the LO frequency, while $\Delta\nu_{\text{PL}}$, can be utilized for the direct stabilization of the laser frequency to the clock cell, that is the goal of this work.  

It is interesting to note that $\Delta\nu_{\text{LO}}$ is the error of the local oscillator already corrected for the light shift, showing significant analogies with \cite{Yudin2018a}. This would suggest that the stabilization of the pumping laser is not strictly required, at least from a theoretical point of view. However, such stabilization is necessary due to the limited precision in the knowledge of the coefficients, their variation in time and with the experimental conditions, and the fact that \cref{eq:system} represents a first-order approximation. By stabilizing the laser frequency, the light shift correction is enforced, and the contribution of the second-order terms remains constant, thereby preserving the stability of the clock.

It is worth noting that both parameters are stabilized using information derived from the same atoms involved in the clock. This is achieved through a more complex clock sequence and the ability to process and extract information in real-time, requiring significant and dedicated processing capability. If necessary, the information processing can be simplified by bypassing the matrix-inversion step. Specifically, when $\frac{B_1}{D_1} \ll \frac{B_2}{D_2}$, $e_1$ can be directly used to stabilize the LO frequency, effectively reducing $\Delta\nu_{\text{LO}}$ close to zero. In such conditions, $e_2$ becomes almost exclusively dependent on the pumping-laser frequency error and can be used directly for its stabilization, thereby bringing the frequency error close to zero and further aligning the LO with the unperturbed atomic frequency $\nu_{21}$. By iterating this process, in steady state, the LO and the PL will tend to $\nu_{21}$ and $\nu_{32}$ respectively.  However, it is important to note that, in this context, the sequence 2 is perceived as a dead time, leading to an increase in the Dick effect that requires careful evaluation and management.

\begin{figure*}[ht]
    \centering
    \includegraphics[width=0.8\textwidth]{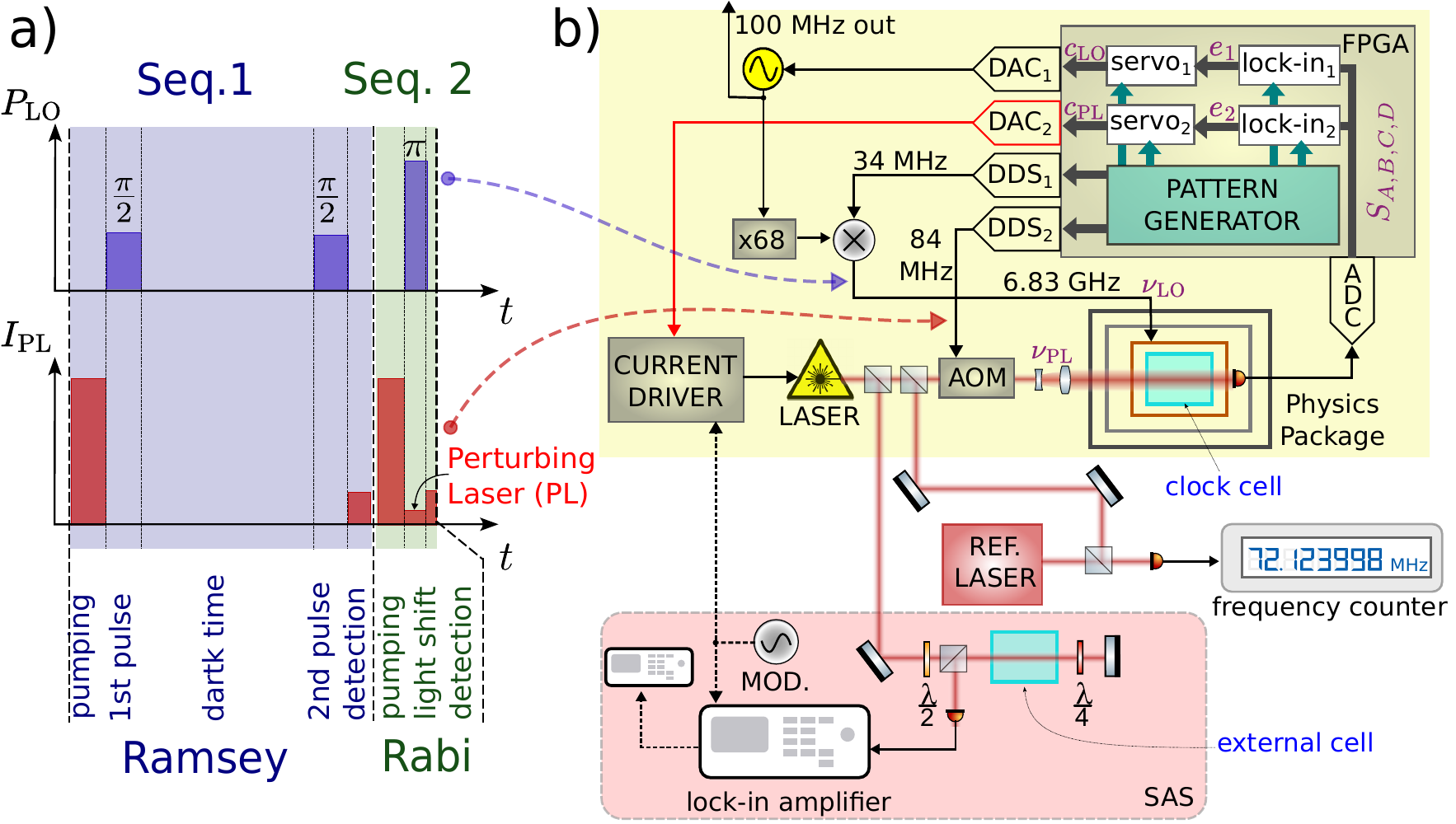}
    \caption{Implementation of the LSLL technique in a pulsed optically pumped clock: a) Ramsey and Rabi sequences with related steps. b) Experimental setup. In the highlighted upper inset, the clock setup is shown, consisting of the FPGA-based electronics package at the top and the optics and physics packages at the inset bottom. The laser is locked to the clock cell and then compared to an external reference laser for its frequency measurement and to the external cell, which now, no longer being necessary, serves as a monitor to display the signal obtained with saturated absorption spectroscopy, for comparison.}
    \label{fig:setup}
\end{figure*}

\section{Implementation and Results}
\label{sec:results}
The technique detailed in Section \ref{sec:technique} was implemented and verified in a Rubidium POP clock. The clock setup has been extensively described in \cite{Gozzelino2023}, \cite{Micalizio2021} for what concerns recent results and in \cite{Eln} and \cite{MTT17} for what concerns the FPGA-based electronics framework. Here, we show in \cref{fig:setup} a simplified diagram of the experimental setup, along with a brief description of it. 

Atoms are prepared using a DFB laser, which is pulsed by means of an acousto-optic modulator (AOM). The atomic sample is then interrogated using a microwave signal, which is derived from a ultra-stable oscillator. The atomic signal is obtained detecting the transmission of a weak laser pulse with a photodiode and acquiring it with a Field Programmable Gate Array (FPGA) through an analog-to-digital converter (ADC).

The clock's pulsed operation is ensured by an embedded pattern generator, which manages the signal processing, as well as the amplitude and frequency modulations of the direct digital synthesizers (DDSs). 
The DDSs allow us to modulate both the microwave and the laser. The laser frequency is monitored using an external buffer-gas-free cell, through saturated absorption spectroscopy (SAS), and with an optical beat-note with an external reference laser. The clock is, in turn, measured with respect to an active hydrogen maser, which has better stability and frequency drift compared to the clock under test.

In this setup, the LSLL technique stabilizes the laser by processing the atomic signal and acting on the driving current of the laser diode. The atomic signal is processed by two lock-in amplifiers to obtain the two error signals, $e_1$ and $e_2$, which are directly fed into two pure integrative controllers that calculate $c_{\text{LO}}$ and $c_{\text{PL}}$, the corrections to be applied to the tuning voltage of the local oscillator (LO) and the laser's supply current through dedicated digital to analog converters (DACs) (see \cref{fig:setup}). The inversion of the matrix is not implemented, for simplicity. 
Thanks to this technique, the laser is maintained automatically in resonance with the $\ket{F_g=1} \rightarrow \ket{F_e=0,1,2}$ D$_2$ transition, shifted by the buffer gas. The collisional broadening of the buffer gas leads to a $\Gamma_3$ of $\simeq\SI{1}{\giga\hertz}$, much larger than the hyperfine splitting of the excited state, leading to a single dispersive signal \cite{Affolderbach2005}. The driving RF frequency of the AOM, now $\SI{-84}{\mega\hertz}$, can take on an absolutely arbitrary value, as it is no longer necessary to bridge the resonance frequency of the atoms in the clock cell (shifted by the buffer gas) to the resonance frequency of an external reference cell (without buffer gas).

The amplitude of the microwave field is stabilized using the technique detailed in \cite{Gozzelino2018}, while, for simplicity, the laser power is stabilized using a low-bandwidth digital control, without reducing the fast laser intensity noise.

A first demonstration of the LSLL technique was already presented in 2017 \cite{MTT17}. There, the low-light-shift POP configuration was alternated with the high-light-shift continuous double resonance (CWDR) one. Although the results were encouraging, the CWDR interrogation quickly proved to be suboptimal, as it consisted of a single step where the atoms simultaneously interacted with laser and microwave. In this work, we offer a more comprehensive examination of the technique by utilizing the Rabi interrogation, as opposed to the previous method. The Rabi sequence offers three degrees of freedom, enabling independent optimization of atom number, light shift amount, and detection sensitivity, resulting in more effective and improved outcomes.

\begin{figure}[ht]
    \centering
    \includegraphics[width=\columnwidth]{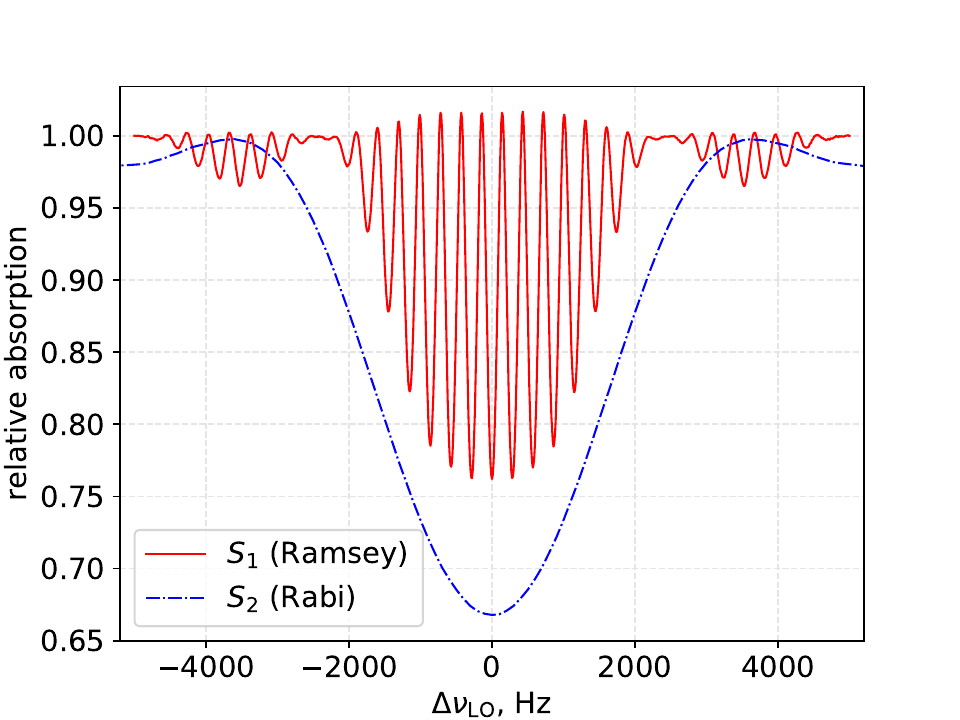}
    \caption{Red continuous trace: Measured absorption signal $S_1$ as a function of the microwave detuning $\Delta\nu_{\text{LO}}$, obtained with the  Ramsey sequence ($T= \SI{3}{\milli\second}$). Blue line-dotted trace: absorption signal $S_2$ obtained with Rabi sequence (pulse length \SI{0.27}{\milli\second}). }
    \label{fig:lines}
\end{figure}

As shown in \cref{fig:setup}a, in the present implementation a Ramsey and a Rabi sequences are alternated. The first one is composed of a short optical pumping pulse (\SI{0.4}{\milli\second}), followed by microwave Ramsey sequence (two $\pi/2$ pulses \SI{0.4}{\milli\second} long, separated by a free-evolution time $T=\SI{3}{\milli\second}$). Finally, a weak probe pulse of duration $\tau_{d1}=\SI{0.3}{\milli\second}$, is sent for measuring the absorption of the atomic vapor. The spectroscopic signal $S_1$ is obtained by integrating the photodiode signal over $\tau_{d1}$. The signal $S_1$ measured as a function of the local oscillator (LO) frequency is shown in \cref{fig:lines} (red trace), where we recognize the characteristic Ramsey fringes in the absorption signal. The sequence is repeated twice to enable a square-wave modulation of the microwave around the central Ramsey fringe (modulation depth ${\nu_{m1} = \SI{80}{\hertz})}$. Following a full modulation cycle, an error signal ${e_1 = S_1(\nu_{\text{LO}}+\nu_{m1}) - S_1(\nu_{\text{LO}}-\nu_{m1})= S_C-S_A }$ is used to perform a bottom-of-the-fringe stabilization of the LO.

The Rabi sequence is composed of a pumping pulse that resets the atomic coherence, i.e. the memory of the atoms, and restores the clock-states population imbalance; then a Rabi pulse is sent, and a small amount of laser power is added to induce a controlled amount of light-shift on the clock transition; at the end, the $\ket{F_g=1}$ clock-state population is probed by integrating the absorption signal during an averaging window of length $\tau_{d2}$, obtaining the signal $S_2$. To keep the dead time of the main clock sequence at a minimum, the Rabi sequence is kept as short as possible (pumping time $\SI{0.4}{\milli\second}$, Rabi pulse $\SI{0.27}{\milli\second}$, $\tau_{d2}=\SI{0.2}{\milli\second}$). A typical spectroscopic signal obtained with this ancillary clock sequence is shown in \cref{fig:lines} (line-dotted blue trace). Also in this case, the LO frequency is modulated, now at $\pm \nu_{m2} = \pm$\SI{1.8}{\kilo\hertz}, to construct an error signal ${e_2=S_2(\nu_{\text{LO}}+\nu_{m2})-S_2(\nu_{\text{LO}}-\nu_{m2})=S_D - S_B}$. 

\begin{figure}[h]
    \centering
    \includegraphics[width=\columnwidth]{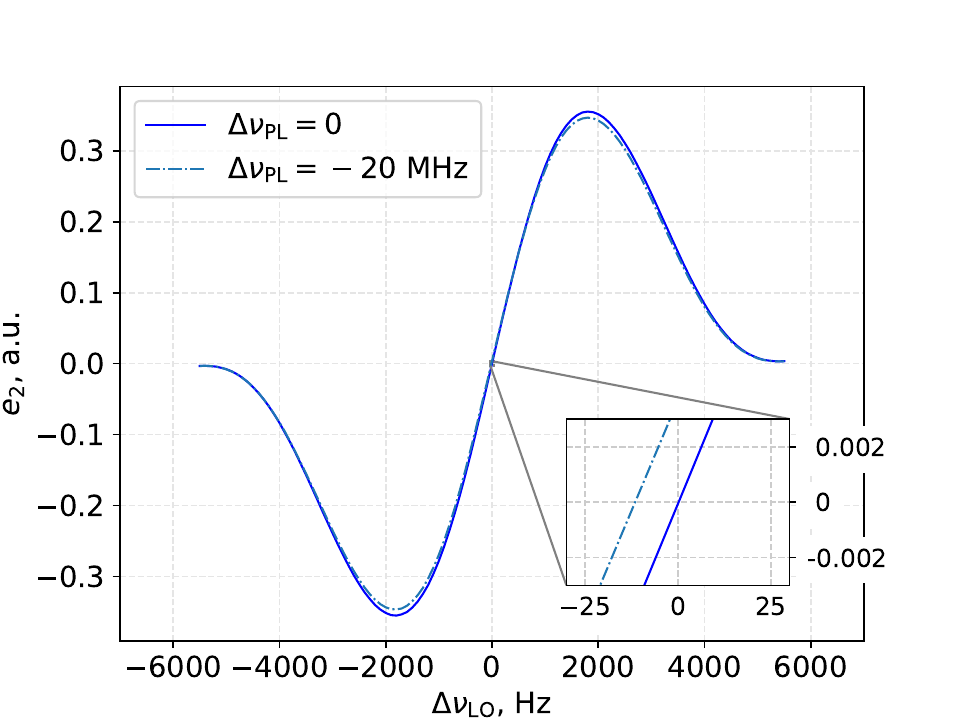}
    \caption{LSLL error signal $e_2$ as a function of the local oscillator detuning $\Delta\nu_{\text{LO}}$. The inset highlights the light shift induced by the pumping-laser detuning $\Delta\nu_{\text{PL}}$. Rabi cycle time: \SI{0.87}{\milli\second} (\SI{0.4}{\milli\second} pumping time, \SI{0.27}{\milli\second} $\pi$-pulse, $\tau_{d2}=\SI{0.2}{\milli\second}$). $I_{\text{PL}}=\SI{3.5e-2}{\milli\watt\per\centi\meter\squared}$. }
    \label{fig:Discriminator}
\end{figure}

In \cref{fig:Discriminator}, we have measured the error signal $e_2$ as a function of the LO frequency at two different pumping-laser detunings: zero in the blue trace  and  \SI{-20}{\mega\hertz} in the case of the green line-dotted trace. 
In the inset, we can clearly observe a \SI{11.5}{\hertz} shift, corresponding to a sensitivity $\beta_2 \simeq  \SI{8.4e-11}{\per\mega\hertz}$. 

\begin{figure}[ht]
    \centering
    \includegraphics[width=\columnwidth]{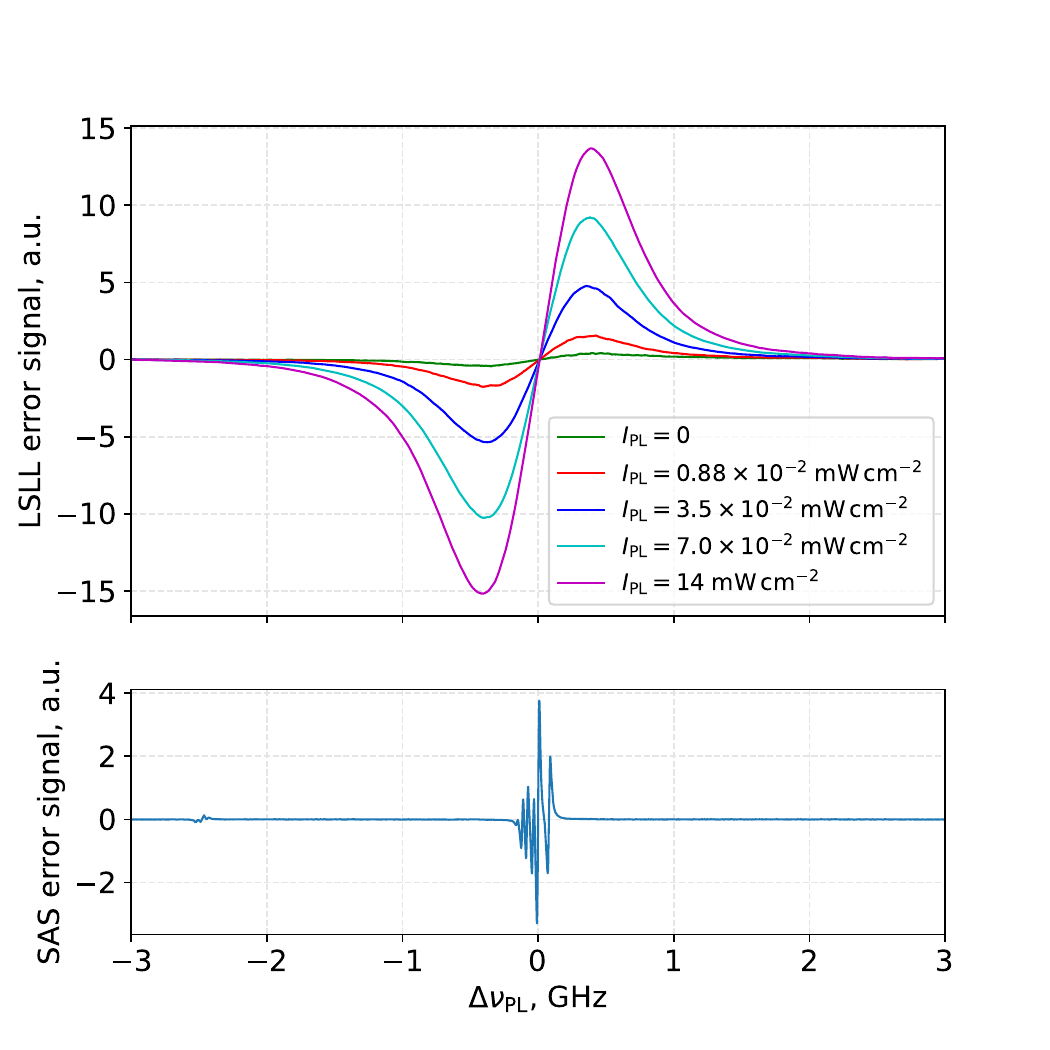}
    \caption{Upper plot: Error signal $e_2$ of the interleaved Rabi clock at fixed microwave detuning ($\Delta\nu_{\text{LO}}\simeq0$) as a function of the laser detuning $\Delta\nu_{\text{PL}}$. Lower plot: spectroscopy signal obtained in a conventional sub-Doppler saturated absorption setup on an external cell filled with Rb (no buffer gas).}
    \label{fig:diff_P}
\end{figure}

In \cref{fig:diff_P}, the error signal $e_2$ is plotted as a function of the laser frequency for different intensities of the perturbing laser $I_{\text{PL}}$. As expected, the light-shift scales linearly with the laser intensity, at least for low perturbations, and the discriminant slope increases accordingly. Higher PL intensity increases the signal and makes the system more robust against technical noise and offsets. However, the perturbing light should be kept low enough to limit excess optical pumping and loss of coherence in the clock states. For our experimental apparatus, we found that an intensity $I_{\text{PL}}$ in the range of \SI{2e-2}{\milli\watt\per\centi\meter\squared} and \SI{4e-2}{\milli\watt\per\centi\meter\squared} is a good compromise for having a strong enough signal with minimal perturbation to the system. With $I_{\text{PL}}=\SI{3.5e-2}{\milli\watt\per\centi\meter\squared}$, we obtain the following coefficients: $D_1\simeq \SI{2e4}{\per\hertz}$, $B_1\simeq \SI{0.3}{\per\mega\hertz}$, $D_2\simeq \SI{2e3}{\per\hertz}$, $B_2\simeq \SI{400}{\per\mega\hertz}$.

In \cref{fig:diff_P} (lower plot), we also reported for comparison the familiar error signal obtained on a saturated absorption spectroscopy setup, utilizing an isotopically enriched $\mathrm{^{87}Rb}$ cell with no buffer gas. The SAS signal is Doppler-free, leading to much narrower resonances. However, the available signal is typically small, requiring fast modulation and lock-in detection, and the capture range is limited to a few \si{\mega\hertz}. With the LSLL technique, the error signal has more than \SI{2}{\giga\hertz} capture range and only one resonance, making the stabilization much more robust and appealing for industrial applications and unsupervised operation.

\subsection{Laser and clock stabilities}
\label{sec:stabilities}
\begin{figure}[ht]
    \centering
    \includegraphics[width=\columnwidth]{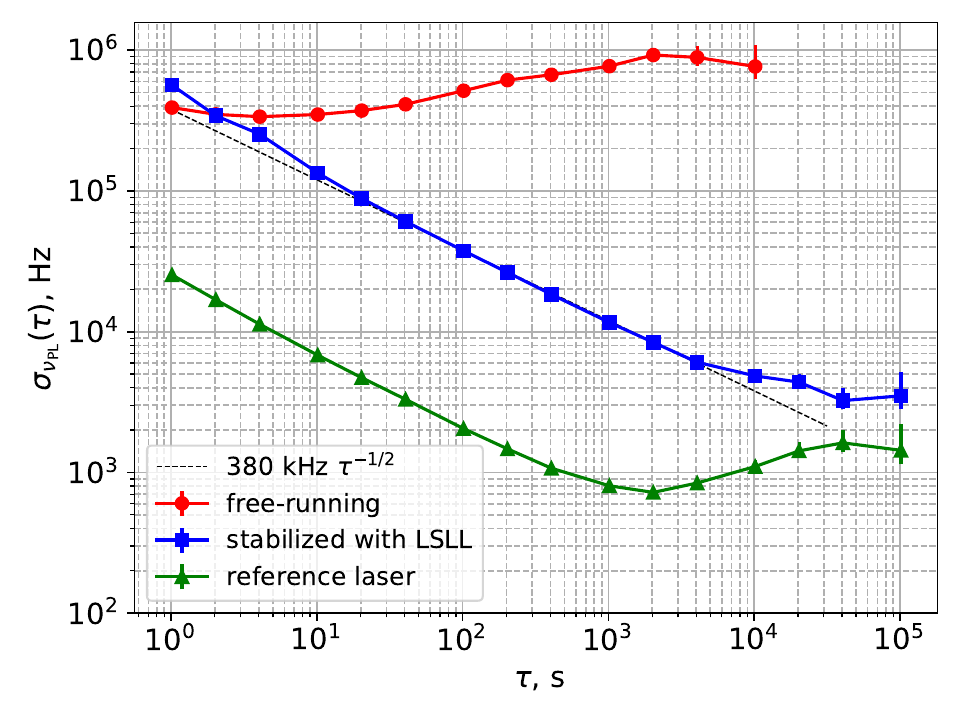}
    \caption{Typical stability of the DFB laser frequency. Red dots: free-running frequency stability (constant current and constant temperature, no active stabilization). Blue squares: laser stabilized with the LSLL technique. Green triangles: stability of the reference laser used for characterization.}
    \label{fig:stab_nuL}
\end{figure}

In this section we report the typical laser-frequency stability that we obtained with the LSLL technique. In \cref{fig:stab_nuL}, the free-running stability of our laser diode is reported in terms of Allan deviation (red dots). The laser frequency stability was measured with respect to a DFB laser of the same class stabilized on a SAS setup. An upper limit of the frequency stability of the reference laser is reported in green dots. The latter was obtained measuring the frequency stability of two identical SAS setups in a dedicated independent measurement. With blue squares, we report the frequency stability obtained with the LSLL technique.   
The short term stability is limited by the detection noise and the locking time constant can be as short as 100 ms (a few clock cycles). The noise is white up to $\tau=\SI{4000}{\second}$, and reaches  \SI{3.5}{\kilo\hertz} for an averaging time of one day. This is in line with the results obtained in tabletop spectroscopy setups utilizing much narrower sub-Doppler atomic resonance as a reference. Indeed, for long averaging times the reference laser that we used to perform the measurement has itself a stability of a few kilohertz for daily timescales, and we cannot exclude that  it is itself contributing to the observed instability.

\begin{figure}[ht]
    \centering
    \includegraphics[width=\columnwidth]{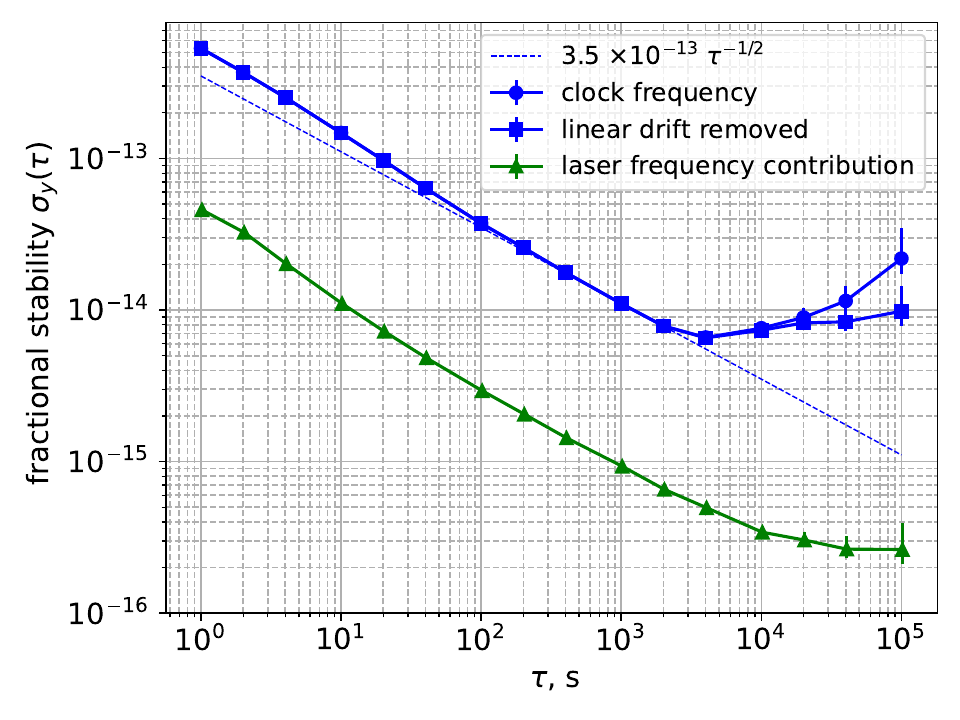}
    \caption{The stability of the POP clock, obtained using the LSLL technique (blue dots), is shown in terms of Allan deviation. The stability of the clock with a linear drift removed (\SI{2.4e-14}{\per\day}) is also depicted with blue squares. The LSLL contribution to the clock stability is shown in green squares, calculated by measuring the laser frequency stability on a dedicated setup and multiplying it by the sensitivity of the clock (\SI{-1e-13}{\per\mega\hertz}). The measurement bandwith is \SI{0.5}{\hertz}. }
    \label{fig:stab_clock}
\end{figure}

In \cref{fig:stab_clock}, we report a typical stability obtained with our POP clock  with the laser stabilized on the clock cell with the LSLL method. This stability exhibits white frequency noise of $\SI{3.2e-13} \tau^{-1/2}$, capable of reaching \SI{1e-14} at \SI{1000}{s} and maintaining it up to \SI{100000}{s} after removing the drift. The technique slightly increases the dead time, from \SI{24.5}{\percent} to \SI{36.7}{\percent}. 
However, the Dick effect's short-term impact remains negligible, increasing from $\sigma_y(1 \, \mathrm{s}) = \num{2e-14}$ to  
\num{3e-14}  [31]. We achieved the same short-term stability even without interleaving the second sequence, indicating that the observed stability is due to other causes, such as laser intensity noise.

In green, we report the contribution to the clock stability resulting from fluctuations of the laser frequency. This contribution is determined by multiplying the pumping-laser frequency stability, measured synchronously using a beatnote setup against the reference laser, by the clock sensitivity $\beta =\SI{-1e-13}{\per\mega\hertz}$. It is evident that the contribution arising from the laser frequency is more than one order of magnitude lower than the measured clock stability. Indeed, the POP clock stability reaches a flicker floor that is consistent with the most recent results obtained with a laser stabilized on a saturated absorption setup \cite{Gozzelino2023}. 

\section{Conclusions}
The proposed Light-Shift Laser-Lock (LSLL) technique represents a promising advancement in simplifying the laser setup of compact atomic clocks while mitigating the light-shift effects at the same time. 
The method is based on a differential measurement, in which a controlled amount of resonant light is purposely introduced on a dedicated interleaved clock sequence. 
In this way, the shift on the clock transition is probed and used as a discriminator to stabilize the laser frequency. 

A major advantage of this scheme lies in avoiding  an external reference, as it makes use of the atomic signal already available from the clock interrogation. In this way, the light shift induced by the pump laser is measured and compensated by directly referring to the atoms participating in the clock, putting them in a less perturbed state and reducing the effects of light shift on the clock stability.

The LSLL technique  offers a wide capture range when used to stabilize lasers resonant to strong allowed transitions, making it advantageous for automated operation. Unlike other methods, The LSLL does not require frequency modulation, eliminating an additional source of clock instability \cite{Huang2023}.

This method has shown particular promise for high-performance vapor-cell clocks, as demonstrated in a Rb POP clock. There, it provides an error signal with a single zero-crossing and a capture range exceeding \SI{2}{\giga\hertz}. The frequency stability of the laser is below 5 kHz at one-day averaging time and is comparable to more conventional laser stabilization techniques based on Doppler-free spectroscopy. With the LSLL, the clock stability reached low $10^{-13}$ at \SI{1}{\second}, with a floor below \SI{1e-14} up to \SI{100000}{s}, a level compliant with the frequency stability specifications of the Galileo second generation onboard clocks. These results, combined with hardware simplification, make LSLL advantageous for industrial and space applications where compact and robust automatic laser frequency stabilization is highly valued. 



Overall, the technique holds potential benefits for miniaturized clocks \cite{Batori2022, Kitching2018, u-cell_22}, as it requires little or no additional hardware, thus maintaining small size, weight and power consumption (SWaP) while potentially improving long-term frequency stability.  Moreover, even if demonstrated on a vapor-cell clock, the LSLL is rather general and could be adapted also for other atomic clock technologies involving a 3-level system, such as compact microwave ion clocks  or even compact optical frequency standards \cite{Schwindt2018, Manai2020}. 

\begin{acknowledgments}
We thank the Atomic Clocks Team of Leonardo S.p.A. and the European Space Agency (ESA) for making the Physics Package Engineering Model ’EM2 PP’ used for the clock measurements available. EM2 PP is ESA property and it was developed in the frame of the General Support Programme (GSTP), ESA Contract 4000118182/16/NL/GLC/FK. We also thank M. Barbiero for careful reading of the manuscript and useful suggestions.
\end{acknowledgments}


\bibliography{biblio}

\end{document}